\documentclass[conference,letterpaper]{IEEEtran}

\usepackage{cite}
\usepackage{latexsym}
\usepackage{amsmath, amssymb}
\usepackage{mathrsfs}
\usepackage[dvips,dvipdf]{graphicx}
\usepackage{subfigure}
\usepackage{algorithm}
\usepackage{algorithmic}

\newcommand{\A}{\mathbf{A}}
\newcommand{\B}{\mathbf{B}}
\newcommand{\C}{\mathbf{C}}
\newcommand{\D}{\mathbf{D}}

\newcommand{\G}{\mathbf{G}}
\newcommand{\I}{\mathbf{I}}

\newcommand{\Q}{\mathbf{Q}}

\newcommand{\s}{\mathbf{s}}

\newcommand{\U}{\mathbf{U}}

\newcommand{\X}{\mathbf{X}}
\newcommand{\x}{\mathbf{x}}

\newcommand{\y}{\mathbf{y}}
\newcommand{\GG}{\mathbf{\Gamma}}

\newcommand{\Hc}{\mathbf{H}}

\newcommand{\mLambda}{\mathbf{\Lambda}}

\newcommand{\tr}{\mathrm{Tr}}

\newcommand{\norm}[1]{\mbox{$\left\lVert #1 \right\rVert$}}

\newcommand{\Pwrcon}{\Omega_{+}(P)}

\newtheorem{thm}{Theorem}

\newcommand{\mtrx}[1]{\mbox{$\left[\begin{array} #1 \end{array}\right]$}}
\newcommand{\diag}[1]{\mbox{Diag}\mbox{$\left\{ #1 \right\}$}}

\begin{document}

\title{Conjugate Gradient Projection Approach for Multi-Antenna Gaussian Broadcast Channels}

\author{
\authorblockN{Jia Liu\authorrefmark{1}, Y. Thomas Hou\authorrefmark{1}, and Hanif D. Sherali\authorrefmark{2}}
\authorblockA{\authorrefmark{1} Department of Electrical and Computer Engineering\\
\authorrefmark{2} Department of Industrial and Systems Engineering \\
Virginia Polytechnic Institute and State University, Blacksburg, VA 24061
\\Email: \{kevinlau, thou, hanifs\}@vt.edu}
 }

\maketitle

\begin{abstract}
THIS PAPER IS ELIGIBLE FOR THE STUDENT PAPER AWARD. It has been shown recently that the dirty-paper
coding is the optimal strategy for maximizing the sum rate of multiple-input multiple-output
Gaussian broadcast channels (MIMO BC). Moreover, by the channel duality, the nonconvex MIMO BC sum
rate problem can be transformed to the convex dual MIMO multiple-access channel (MIMO MAC) problem
with a sum power constraint. In this paper, we design an efficient algorithm based on conjugate
gradient projection (CGP) to solve the MIMO BC maximum sum rate problem. Our proposed CGP algorithm
solves the dual sum power MAC problem by utilizing the powerful concept of Hessian conjugacy. We
also develop a rigorous algorithm to solve the projection problem. We show that CGP enjoys provable
convergence, nice scalability, and great efficiency for large MIMO BC systems.
\end{abstract}

\section{Introduction}
Recently, researchers have shown great interests in characterizing the capacity region for
multiple-input multiple-output broadcast channels (MIMO BC) and MIMO multiple-access channels (MIMO
MAC). In particular, although the general capacity region for MIMO BC remains an open problem
\cite{Goldsmith03:MIMO_Limit}, the sum rate region has been shown achievable by the dirty-paper
coding strategy \cite{Caire03:DPC, Costa03:DPC}. Moreover, by the remarkable channel {\em duality}
between MIMO BC and MIMO MAC established in \cite{Vishwanath03:duality, Viswanath03:duality,
Yu06:duality}, the nonconvex MIMO BC sum rate problem can be transformed to the convex dual MIMO
MAC problem with a sum power constraint.

However, although the standard interior point convex optimization method can be used to solve the
sum power MIMO MAC problem, its complexity is considerably higher than those methods that exploit
the special structure of the sum power MIMO MAC problem. Such specifically designed algorithms
include the minimax method (MM) by Lan and Yu \cite{Lan_Yu_GLOBECOM04}, the steepest descent (SD)
method by Viswanathan {\em et al.} \cite{Viswanathan03:MIMO_BC_SD}, the dual decomposition (DD)
method by Yu \cite{Yu_CISS03}, and two iterative water-filling methods (IWFs) by Jindal {\em et
al.}\cite{Jindal04:MIMO_BC_IWF}. Among these algorithms, MM is more complex than the others having
the linear complexity. SD and DD have longer running time per iteration than IWFs due to line
searches and the inner optimization, respectively. Both IWFs in \cite{Jindal04:MIMO_BC_IWF},
however, do not scale well as the number of users, denoted by $K$, increases. The reason is that in
each iteration of IWFs, the most recently updated solution only accounts for a fraction of $1/K$ in
the effective channels' computation. The authors of \cite{Jindal04:MIMO_BC_IWF} proposed a hybrid
algorithm as a remedy. But the hybrid algorithm introduces additional complexity in implementation
and its performance depends upon the empirical switch timing, which, in turn, depends upon specific
problems. In addition, one of the IWFs in \cite{Jindal04:MIMO_BC_IWF}, although converges
relatively faster than the other one, requires a total storage size for $K^{2}$ input covariance
matrices. These limitations of the existing algorithms motivate us to design an efficient and
scalable algorithm with a modest storage requirement for solving large MIMO BC systems.

Our major contribution in this paper is that we design a fast algorithm based on {\em Conjugate
Gradient Projection} (CGP) approach. Our algorithm is inspired by \cite{Ye03:MIMO_SH_AdHoc}, where
a gradient projection method was used to heuristically solve another nonconvex maximum sum rate
problem for single-hop MIMO-based ad hoc networks with mutual interference. However, unlike
\cite{Ye03:MIMO_SH_AdHoc}, we use the {\em conjugate} gradient directions instead of gradient
directions to eliminate the ``zigzagging'' phenomenon encountered in \cite{Ye03:MIMO_SH_AdHoc}.
Also, we develop a rigorous algorithm to exactly solve the projection problem (in
\cite{Ye03:MIMO_SH_AdHoc}, the way of handling gradient projection is based on heuristic: The
authors simply set the first derivative to zero to get the solution when solving the constrained
Lagrangian dual of the projection problem). The attractive features of our proposed CGP are as
follows:
\begin{enumerate}
\item CGP is extremely fast, and enjoys provable convergence as well as nice scalability. As opposed to IWFs, the number of iterations required for convergence in CGP is
very insensitive to the increase of the number of users.

\item CGP has the desirable linear complexity. By adopting the inexact line search
method called ``Armijo's Rule'', we show that CGP has a comparable complexity to IWFs per
iteration, and requires much fewer iterations for convergence in large MIMO BC systems.

\item CGP has a modest memory requirement: It only needs the solution information from the previous step, as opposed to one of the IWFs, which
requires the solution information from previous $K-1$ steps. Moreover, CGP is very intuitive and
easy to implement.
\end{enumerate}

The remainder of this paper is organized as follows. In Section~\ref{sec:model}, we discuss the
network model and formulation. Section~\ref{sec:framework} introduces the key components in our CGP
framework, including conjugate gradient computation and how to perform projection. We analyze and
compare the complexity of CGP with other existing algorithms in Section~\ref{sec:complexity}.
Numerical results are presented in Section~\ref{sec:results}. Section~\ref{sec:conclusions}
concludes this paper.

\section{System Model and Problem Formulation} \label{sec:model}
We first introduce notation. We use boldface to denote matrices and vectors. For a complex-valued
matrix $\A$, $\A^{*}$ and $\A^{\dag}$ denotes the conjugate and conjugate transpose of $\A$,
respectively. $\tr\{\A\}$ denotes the trace of $\A$. We let $\I$ denote the identity matrix with
dimension determined from context. $\A \succeq 0$ represents that $\A$ is Hermitian and positive
semidefinite (PSD). $\diag{\A_{1} \ldots \A_{n}}$ represents the block diagonal matrix with
matrices $\A_{1},\ldots,\A_{n}$ on its main diagonal.

Suppose that a MIMO Gaussian broadcast channel has $K$ users, each of which is equipped with
$n_{r}$ antennas, and the transmitter has $n_{t}$ antennas. The channel matrix for user $i$ is
denoted as $\Hc_{i} \in \mathbb{C}^{n_{r} \times n_{t}}$.

In \cite{Caire03:DPC, Vishwanath03:duality, Viswanath03:duality, Yu06:duality}, it has been shown
that the maximum sum rate capacity of MIMO BC is equal to the dirty-paper coding region, which can
be computed by solving the optimization problem as follows:
\begin{equation} \label{eqn_bc_sumrate}
\begin{array}{rl}
\mbox{Maximize} & \sum_{i=1}^{K} \log \frac{\det \left(\I + \Hc_{i} \left(\sum_{j=1}^{i} \GG_{j}
\right) \Hc_{i}^{\dag}) \right)}{\det \left(\I + \Hc_{i} \left( \sum_{j=1}^{i-1} \GG_{j} \right)
\Hc_{i}^{\dag}\right) }
\\
\mbox{subject to} & \GG_{i} \succeq 0, \quad i=1,2,\ldots,K \\
& \sum_{i=1}^{K} \tr (\GG_{i}) \leq P,
\end{array}
\end{equation}
where $\GG_{i} \in \mathbb{C}^{n_{t}\times n_{t}}$, $i=1,\ldots,K$, are the downlink input
covariance matrices. It is evident that (\ref{eqn_bc_sumrate}) is a nonconvex optimization problem.
However, the authors in \cite{Vishwanath03:duality, Yu06:duality} showed that due to the duality
between MIMO BC and MIMO MAC, (\ref{eqn_bc_sumrate}) is equivalent to the following MIMO MAC
problem with a sum power constraint:
\begin{equation} \label{eqn_mac_equiv}
\begin{array}{rl}
\mbox{Maximize} & \log \det \left( \I + \sum_{i=1}^{K} \Hc_{i}^{\dag} \Q_{i} \Hc_{i} \right) \\
\mbox{subject to} & \Q_{i} \succeq 0, \quad i=1,2,\ldots,K\\
& \sum_{i=1}^{K} \tr (\Q_{i}) \leq P, \\
\end{array}
\end{equation}
where $\Q_{i} \in \mathbb{C}^{n_{r}\times n_{r}}$, $i=1,\ldots,K$ are the uplink input covariance
matrices. For convenience, we use the matrix $\Q = \mtrx{{cccc} \Q_{1} & \Q_{2} & \ldots & \Q_{K}}$
to denote the set of all uplink input covariance matrices, and let $F(\Q) = \log \det \left( \I +
\sum_{i=1}^{K} \Hc_{i}^{\dag} \Q_{i} \Hc_{i} \right)$ represent the objective function of
(\ref{eqn_mac_equiv}). After solving (\ref{eqn_mac_equiv}), we can recover the solutions of
(\ref{eqn_bc_sumrate}) by the mapping proposed in \cite{Vishwanath03:duality}.

\section{Conjugate Gradient Projection for MIMO BC} \label{sec:framework}
In this paper, we propose an efficient algorithm based on conjugate gradient projection (CGP) to
solve (\ref{eqn_mac_equiv}). CGP utilizes the important and powerful concept of Hessian conjugacy
to deflect the gradient direction so as to achieve the superlinear convergence rate
\cite{Bazaraa_Sherali_Shetty_93:NLP} similar to that of the well-known quasi-Newton methods (e.g.,
BFGS method). Also, gradient projection is a classical method originally proposed by Rosen
\cite{Rosen60:GrdPrj} aiming at solving constrained nonlinear programming problems. But its
convergence proof has not been established until very recently
\cite{Bazaraa_Sherali_Shetty_93:NLP}. The framework of CGP for solving (\ref{eqn_mac_equiv}) is
shown in Algorithm~\ref{alg_mgp}.
\begin{algorithm}
{\footnotesize \caption{Gradient Projection Method} \label{alg_mgp}
\begin{algorithmic}
\STATE {\bf Initialization:} \\
\STATE \quad Choose the initial conditions $\Q^{(0)} = [ \Q_{1}^{(0)}, \Q_{2}^{(0)}, \ldots, \Q_{K}^{(0)}]^{T}$. Let \\
\STATE \quad $k=0$. \\
\STATE {\bf Main Loop:} \\
\STATE \quad 1. Calculate the conjugate gradients $\G_{i}^{(k)}$, $i= 1,2, \ldots, K$. \\
\STATE \quad 2. Choose an appropriate step size $s_{k}$. Let $\Q_{i}^{'(k)} = \Q_{i}^{(k)} + s_{k}
\G_{i}^{(k)}$, \\
\STATE \quad \quad for $i=1,2,\ldots,K$. \\
\STATE \quad 3. Let $\bar{\Q}^{(k)}$ be the projection of $\Q^{'(k)}$ onto $\Pwrcon$, where
$\Pwrcon \triangleq $\\
\STATE \quad \quad $\{ \Q_{i}, \,\, i=1,\ldots,K | \Q_{i} \succeq 0, \sum_{i=1}^{K} \tr\{ \Q_{i} \} \leq P \}$. \\
\STATE \quad 4. Choose appropriate step size $\alpha_{k}$. Let $\Q_{l}^{(k+1)} = \Q_{l}^{(k)} +
\alpha_{k}(\bar{\Q}_{i}^{(k)} - $ \\
\STATE \quad \quad $\Q_{i}^{(k)})$, $i=1,2,\ldots,K$. \\
\STATE \quad 5. $k=k+1$. If the maximum absolute value of the elements in $\Q_{i}^{(k)} - $\\
\STATE \quad \quad $\Q_{i}^{(k-1)}< \epsilon$, for $i=1,2,\ldots,L$, then stop; else go to step 1.
\end{algorithmic}}
\end{algorithm}

Due to the complexity of the objective function in (\ref{eqn_mac_equiv}), we adopt the inexact line
search method called ``Armijo's Rule'' to avoid excessive objective function evaluations, while
still enjoying provable convergence \cite{Bazaraa_Sherali_Shetty_93:NLP}. The basic idea of
Armijo's Rule is that at each step of the line search, we sacrifice accuracy for efficiency as long
as we have sufficient improvement. According to Armijo's Rule, in the $k^{th}$ iteration, we choose
$\sigma_{k}=1$ and $\alpha_{k} = \beta^{m_{k}}$ (the same as in \cite{Ye03:MIMO_SH_AdHoc}), where
$m_{k}$ is the first non-negative integer $m$ that satisfies
\begin{eqnarray} \label{eqn_Armijo}
\!\!\!\!\!\! && F(\Q^{(k+1)}) - F(\Q^{(k)})\geq \sigma \beta^{m} \langle
\G^{(k)}, \bar{\Q}^{(k)} - \Q^{(k)} \rangle \nonumber\\
\!\!\!\!\!\! && = \sigma \beta^{m} \sum_{i=1}^{K} \tr \left[ \G_{i}^{\dag(k)}
\left(\bar{\Q}_{i}^{(k)} - \Q_{i}^{(k)}\right) \right],
\end{eqnarray}
where $0 < \beta < 1$ and $0 < \sigma < 1$ are fixed scalars.

Next, we will consider two major components in the CGP framework: 1) how to compute the conjugate
gradient direction $\G_{i}$, and 2) how to project $\Q^{'(k)}$ onto the set $\Pwrcon \triangleq \{
\Q_{i}, \,\, i=1,\ldots,K | \Q_{i} \succeq 0, \sum_{i=1}^{K} \tr\{ \Q_{i} \} \leq P \}$.

\subsection{Computing the Conjugate Gradients} The gradient $\bar{\G}_{i} \triangleq \nabla_{\Q_{i}}
F(\Q)$ depends on the partial derivatives of $F(\Q)$ with respect to $\Q_{i}$. By using the formula
$\frac{\partial \ln \det(\A+\B\X\C)}{\partial \X} = \left[ \C(\A+\B\X\C)^{-1}\B \right]^{T}$
\cite{Ye03:MIMO_SH_AdHoc, Magnus_Neudecker99:Mtrx_Diff_Calc}, we can compute the partial derivative
of $F(\Q)$ with respect to $\Q_{i}$ as follows (by letting $\A = \I + \sum_{j=1,j \ne i}^{K}
\Hc_{j}^{\dag} \Q_{j} \Hc_{j}$, $\B = \Hc_{i}^{\dag}$, $\X = \Q_{i}$, and $\C = \Hc_{i}$):
\begin{eqnarray} \label{eqn_partial}
\frac{\partial F(\Q)}{\partial \Q_{i}} &=& \frac{\partial}{\partial \Q_{i}} \log \det \left( \I +
\sum_{j=1}^{K} \Hc_{j}^{\dag} \Q_{j} \Hc_{j} \right) \nonumber \\
&=& \left[ \Hc_{i} \left( \I + \sum_{j=1}^{K} \Hc_{j}^{\dag} \Q_{j} \Hc_{j} \right)^{-1}
\Hc_{j}^{\dag} \right]^{T}.
\end{eqnarray}
Further, from the definition $\nabla_{z} f(z) = 2(\partial f(z)/\partial z)^{*}$
\cite{Haykin96:Adpt_Fltr}, we have
\begin{equation} \label{eqn_gradient}
\bar{\G}_{i} = 2\left( \frac{\partial F(\Q)}{\partial \Q_{i}} \right)^{*} \!\!\! = 2 \Hc_{i} \left(
\I + \sum_{j=1}^{K} \Hc_{j}^{\dag} \Q_{j} \Hc_{j} \right)^{-1} \Hc_{i}^{\dag}.
\end{equation}
Then, the conjugate gradient direction can be computed as $\G_{i}^{(k)} = \bar{\G}_{i}^{(k)} +
\rho_{k} \G_{i}^{(k-1)}$. In this paper, we adopt the Fletcher and Reeves' choice of deflection
\cite {Bazaraa_Sherali_Shetty_93:NLP}. The Fletcher and Reeves' choice of deflection can be
computed as
\begin{equation} \label{eqn_conjuate}
\rho_{k} = \frac{\| \bar{\G}_{i}^{(k)} \|^{2}}{\| \bar{\G}_{i}^{(k-1)} \|^{2}}.
\end{equation}
The purpose of deflecting the gradient using (\ref{eqn_conjuate}) is to find $\G_{i}^{(k)}$, which
is the Hessian-conjugate of $\G_{i}^{(k-1)}$. By doing this, we can eliminate the ``zigzagging''
phenomenon encountered in the conventional gradient projection method, and achieve the superlinear
convergence rate \cite{Bazaraa_Sherali_Shetty_93:NLP} without actually storing the matrix of
Hessian approximation as in quasi-Newton methods.

\subsection{Projection onto $\Pwrcon$} Noting from (\ref{eqn_gradient}) that $\G_{i}$ is
Hermitian. We have that $\Q_{i}^{'(k)} = \Q_{i}^{(k)}+s_{k} \G_{i}^{(k)}$ is Hermitian as well.
Then, the projection problem becomes how to simultaneously project a set of $K$ Hermitian matrices
onto the set $\Pwrcon$, which contains a constraint on sum power for all users. This is different
to \cite{Ye03:MIMO_SH_AdHoc}, where the projection was performed on individual power constraint. In
order to do this, we construct a block diagonal matrix $\D = \diag{\Q_{1} \ldots \Q_{K}} \in
\mathbb{C}^{(K\cdot n_{r})\times(K\cdot n_{r})}$. It is easy to recognize that if $\Q_{i} \in
\Pwrcon$, $i=1,\ldots,K$, we have $\tr(\D) = \sum_{i=1}^{K} \tr \left(\Q_{i}\right) \leq P$, and
$\D \succeq 0$. In this paper, we use Frobenius norm, denoted by $\|\cdot\|_{F}$, as the matrix
distance criterion. Then, the distance between two matrices $\A$ and $\B$ is defined as $\| \A - \B
\|_{F} = \left( \tr\left[ (\A-\B)^{\dag} (\A-\B) \right] \right)^{\frac{1}{2}}$. Thus, given a
block diagonal matrix $\D$, we wish to find a matrix $\tilde{\D} \in \Pwrcon$ such that
$\tilde{\D}$ minimizes $\| \tilde{\D} - \D \|_{F}$. For more convenient algebraic manipulations, we
instead study the following equivalent optimization problem:
\begin{equation} \label{eqn_proj_primal_equiv}
\begin{array}{rl}
\mbox{Minimize} & \frac{1}{2} \| \tilde{\D} - \D \|_{F}^{2} \\
\mbox{subject to} & \tr (\tilde{\D}) \leq P, \,\, \tilde{\D} \succeq 0. \\
\end{array}
\end{equation}
In (\ref{eqn_proj_primal_equiv}), the objective function is convex in $\tilde{\D}$, the constraint
$\tilde{\D} \succeq 0$ represents the convex cone of positive semidefinite matrices, and the
constraint $\tr (\tilde{\D}) \leq P$ is a linear constraint. Thus, the problem is a convex
minimization problem and we can exactly solve this problem by solving its Lagrangian dual problem.
Associating Hermitian matrix $\X$ to the constraint $\tilde{\D} \succeq 0$, $\mu$ to the constraint
$\tr (\tilde{\D}) \leq P$, we can write the Lagrangian as
\begin{eqnarray} \label{eqn_prj_lagrangian}
g(\X, \mu) &=& \min_{\tilde{\D}} \left\{ (1/2) \| \tilde{\D} - \D \|_{F}^{2} -
\tr(\X^{\dag} \tilde{\D}) \right.\nonumber\\
&& \quad \quad \,\,+ \left. \mu \left(\tr(\tilde{\D})-P \right) \right\}.
\end{eqnarray}
Since $g(\X, \mu)$ is an unconstrained convex quadratic minimization problem, we can compute the
minimizer of (\ref{eqn_prj_lagrangian}) by simply setting the derivative of
(\ref{eqn_prj_lagrangian}) (with respect to $\tilde{\D}$) to zero, i.e., $(\tilde{\D} - \D) -
\X^{\dag} + \mu \I = 0$. Noting that $\X^{\dag} = \X$, we have $\tilde{\D} = \D - \mu \I + \X$.
Substituting $\tilde{\D}$ back into (\ref{eqn_prj_lagrangian}), we have
\begin{eqnarray}
&& \!\!\!\!\!\!\!\!\!\!\!\!\!g(\X,\mu) = \frac{1}{2} \norm{ \X - \mu\I }_{F}^{2} - \mu P + \tr \left[ \left(\mu\I - \X \right) \left(\D + \X -\mu\I \right) \right] \nonumber\\
&& \quad \,\, = -\frac{1}{2} \norm{ \D - \mu\I + \X }_{F}^{2} - \mu P + \frac{1}{2} \| \D \|^{2}.
\end{eqnarray}
Therefore, the Lagrangian dual problem can be written as
\begin{equation}\label{eqn_prj_dual}
\begin{array}{rl}
\mbox{Maximize} & -\frac{1}{2} \norm{ \D - \mu\I + \X }_{F}^{2} - \mu P + \frac{1}{2} \| \D \|^{2} \\
\mbox{subject to} & \X \succeq 0, \mu \geq 0.
\end{array}
\end{equation}
After solving (\ref{eqn_prj_dual}), we can have the optimal solution to
(\ref{eqn_proj_primal_equiv}) as:
\begin{equation}
\tilde{\D}^{*} = \D - \mu^{*} \I + \X^{*},
\end{equation}
where $\mu^{*}$ and $\X^{*}$ are the optimal dual solutions to Lagrangian dual problem in
(\ref{eqn_prj_dual}). Although the Lagrangian dual problem in (\ref{eqn_prj_dual}) has a similar
structure as that in the primal problem in (\ref{eqn_proj_primal_equiv}) (having a positive
semidefinitive matrix constraint), we find that the positive semidefinite matrix constraint can
indeed be easily handled. To see this, we first introduce Moreau Decomposition Theorem from convex
analysis.
\begin{thm}\label{thm_moreau}(Moreau Decomposition \cite{Hiriart-Urruty_Lemarechal01:Cnvx_Anl})
Let $\mathcal{K}$ be a closed convex cone. For $\x,\x_{1},\x_{2} \in \mathbb{C}^{p}$, the two
properties below are equivalent:
\begin{enumerate}
\item $\x = \x_{1} + \x_{2}$ with $\x_{1} \in \mathcal{K}$, $\x_{2} \in \mathcal{K}^{o}$ and $\langle \x_{1},\x_{2} \rangle =
0$,
\item $\x_{1} = p_{\mathcal{K}}(\x)$ and $\x_{2} = p_{\mathcal{K}^{o}}(x)$,
\end{enumerate}
where $\mathcal{K}^{o} \triangleq \{ \s \in \mathbb{C}^{p}: \langle \s, \y \rangle \leq 0, \,
\forall \, \y \in \mathcal{K} \}$ is called the polar cone of cone $\mathcal{K}$,
$p_{\mathcal{K}}(\cdot)$ represents the projection onto cone $\mathcal{K}$.
\end{thm}

In fact, the projection onto a cone $\mathcal{K}$ is analogous to the projection onto a subspace.
The only difference is that the orthogonal subspace is replaced by the polar cone.

Now we consider how to project a Hermitian matrix $\A \in \mathbb{C}^{n \times n}$ onto the
positive and negative semidefinite cones. First, we can perform eigenvalue decomposition on $\A$
yielding $\A = \U \diag{\lambda_{i}, \,\, i=1,\ldots,n } \U^{\dag}$, where $\U$ is the unitary
matrix formed by the eigenvectors corresponding to the eigenvalues $\lambda_{i}$, $i=1,\ldots,n$.
Then, we have the positive semidefinite and negative semidefinite projections of $\A$ as follows:
\begin{eqnarray}
\label{eqn_prj_psd} &\A_{+} = \U \diag{ \max\{ \lambda_{i},0\}, i=1,2,\ldots,n } \U^{\dag},& \\
\label{eqn_prj_nsd} &\A_{-} = \U \diag{ \min\{ \lambda_{i},0\}, i=1,2,\ldots,n } \U^{\dag}.&
\end{eqnarray}
The proof of (\ref{eqn_prj_psd}) and (\ref{eqn_prj_nsd}) is a straightforward application of
Theorem~\ref{thm_moreau} by noting that $\A_{+} \succeq 0$, $\A_{-} \preceq 0$, $\langle \A_{+},
\A_{-} \rangle = 0$, $\A_{+} + \A_{-} = \A$, and the positive semidefinite cone and negative
semidefinite cone are polar cones to each other.

We now consider the term $\D - \mu\I + \X$, which is the only term involving $\X$ in the dual
objective function. We can rewrite it as $\D - \mu\I - (-\X)$, where we note that $-\X \preceq 0$.
Finding a negative semidefinite matrix $-\X$ such that $\| \D - \mu\I - (-\X) \|_{F}$ is minimized
is equivalent to finding the projection of $\D - \mu\I$ onto the negative semidefinite cone. From
the previous discussions, we immediately have
\begin{equation} \label{eqn_prj_negX}
-\X = \left(\D - \mu\I \right)_{-}.
\end{equation}
Since $\D - \mu \I = (\D - \mu \I)_{+} + (\D - \mu \I)_{-}$, substituting (\ref{eqn_prj_negX}) back
to the Lagrangian dual objective function, we have
\begin{equation}
\min_{\X} \norm{\D - \mu\I + \X}_{F} = \left(\D - \mu\I \right)_{+}.
\end{equation}
Thus, the matrix variable $\X$ in the Lagrangian dual problem can be removed and the Lagrangian
dual problem can be rewritten to
\begin{equation}\label{eqn_prj_dual_sim}
\begin{array}{rl}
\!\!\!\!\!\!\! \mbox{Maximize} & \!\!\! \psi(\mu) \triangleq -\frac{1}{2} \norm{ \left(\D - \mu\I
\right)_{+} }_{F}^{2} - \mu P + \frac{1}{2} \norm{ \D }^{2} \\
\!\!\!\!\!\!\! \mbox{subject to} & \!\!\! \mu \geq 0.
\end{array}
\end{equation}
Suppose that after performing eigenvalue decomposition on $\D$, we have $\D = \U \mLambda
\U^{\dag}$, where $\mLambda$ is the diagonal matrix formed by the eigenvalues of $\D$, $\U$ is the
unitary matrix formed by the corresponding eigenvectors. Since $\U$ is unitary, we have $\left(\D -
\mu\I \right)_{+} = \U \left(\mLambda - \mu \I \right)_{+} \U^{\dag}$. It then follows that
\begin{equation}
\norm{ \left(\D - \mu\I \right)_{+} }_{F}^{2} = \norm{\left(\mLambda - \mu \I \right)_{+}}_{F}^{2}.
\end{equation}
We denote the eigenvalues in $\mLambda$ by $\lambda_{i}$, $i=1,2,\ldots,K\cdot n_{r}$. Suppose that
we sort them in non-increasing order such that $\mLambda = \diag{\lambda_{1} \,\, \lambda_{2}
\ldots \,\, \lambda_{K\cdot n_{r}}}$, where $\lambda_{1} \geq \ldots \geq \lambda_{K\cdot n_{r}}$.
It then follows that
\begin{equation} \label{eqn_norm}
\norm{\left(\mLambda - \mu \I \right)_{+}}_{F}^{2} = \sum_{j=1}^{K \cdot n_{r}} \left( \max
\left\{0, \lambda_{j} - \mu \right\} \right)^{2}.
\end{equation}
From (\ref{eqn_norm}), we can rewrite $\psi(\mu)$ as
\begin{equation} \label{eqn_psi}
\psi(\mu) = - \frac{1}{2} \sum_{j=1}^{K \cdot n_{r}} \left( \max \left\{0, \lambda_{j} - \mu
\right\} \right)^{2} - \mu P + \frac{1}{2} \norm{\D_{n}}^{2}.
\end{equation}
It is evident from (\ref{eqn_psi}) that $\psi(\mu)$ is continuous and (piece-wise) concave in
$\mu$. Generally, piece-wise concave maximization problems can be solved by using the subgradient
method. However, due to the heuristic nature of its step size selection strategy, subgradient
algorithm usually does not perform well. In fact, by exploiting the special structure,
(\ref{eqn_prj_dual_sim}) can be efficiently solved. We can search the optimal value of $\mu$ as
follows. Let $\hat{I}$ index the pieces of $\psi(\mu)$, $\hat{I}=0,1,\ldots,K\cdot n_{r}$.
Initially we set $\hat{I}=0$ and increase $\hat{I}$ subsequently. Also, we introduce $\lambda_{0} =
\infty$ and $\lambda_{K\cdot n_{r}+1} = -\infty$. We let the endpoint objective value
$\psi_{\hat{I}}\left(\lambda_{0} \right) = 0$, $\phi^{*} = \psi_{\hat{I}} \left( \lambda_{0}
\right)$, and $\mu^{*} = \lambda_{0}$. If $\hat{I} > K \cdot n_{r}$, the search stops. For a
particular index $\hat{I}$, by setting
\begin{equation}
\frac{\partial}{\partial \mu} \psi_{\hat{I}}(\nu) \triangleq \frac{\partial}{\partial \mu} \left(
-\frac{1}{2} \sum_{i=1}^{\hat{I}}\left(\lambda_{i}-\mu \right)^{2} - \mu P \right) = 0,
\end{equation}
we have
\begin{equation}
\mu_{\hat{I}}^{*} = \frac{\sum_{i=1}^{\hat{I}} \lambda_{i} - P }{\hat{I}}.
\end{equation}
Now we consider the following two cases:
\begin{enumerate}
\item If $\mu_{\hat{I}}^{*} \in \left[\lambda_{\hat{I}+1}, \lambda_{\hat{I}}\right] \cap
\mathbb{R}_{+}$, where $\mathbb{R}_{+}$ denotes the set of non-negative real numbers, then we have
found the optimal solution for $\mu$ because $\psi(\mu)$ is concave in $\mu$. Thus, the point
having zero-value first derivative, if exists, must be the unique global maximum solution. Hence,
we can let $\mu^{*}=\mu_{\hat{I}}^{*}$ and the search is done.

\item If $\mu_{\hat{I}}^{*} \notin \left[\lambda_{\hat{I}+1},\lambda_{\hat{I}}\right] \cap \mathbb{R}_{+}$,
we must have that the local maximum in the interval $\left[\lambda_{\hat{I}+1},
\lambda_{\hat{I}}\right] \cap \mathbb{R}_{+}$ is achieved at one of the two endpoints. Note that
the objective value $\psi_{\hat{I}}\left( \lambda_{\hat{I}} \right)$ has been computed in the
previous iteration because from the continuity of the objective function, we have
$\psi_{\hat{I}}\left( \lambda_{\hat{I}} \right) = \psi_{\hat{I}-1}\left( \lambda_{\hat{I}}
\right)$. Thus, we only need to compute the other endpoint objective value
$\psi_{\hat{I}}\left(\lambda_{\hat{I}+1}\right)$. If
$\psi_{\hat{I}}\left(\lambda_{\hat{I}+1}\right) < \psi_{\hat{I}}\left(\lambda_{\hat{I}}\right) =
\phi^{*}$, then we know $\mu^{*}$ is the optimal solution; else let $\mu^{*} =
\lambda_{\hat{I}+1}$, $\phi^{*} = \psi_{\hat{I}}\left( \lambda_{\hat{I}+1} \right)$, $\hat{I} =
\hat{I} + 1$ and continue.
\end{enumerate}
Since there are $K \cdot n_{r}+1$ intervals in total, the search process takes at most $K \cdot
n_{r}+1$ steps to find the optimal solution $\mu^{*}$. Hence, this search is of polynomial-time
complexity $O(n_{r}K)$.

After finding $\mu^{*}$, we can compute $\tilde{\D}^{*}$ as
\begin{equation}
\tilde{\D}^{*} = \left( \D - \mu^{*} \I \right)_{+} = \U \left( \mLambda - \mu^{*} \I \right)_{+}
\U^{\dag}.
\end{equation}
That is, the projection $\tilde{\D}$ can be computed by adjusting the eigenvalues of $\D$ using
$\mu^{*}$ and keeping the eigenvectors unchanged. The projection of $\D_{n}$ onto $\Pwrcon$ is
summarized in Algorithm~\ref{alg_rlt_prj}.
\begin{algorithm}
\caption{Projection onto $\Pwrcon$} \label{alg_rlt_prj}
\begin{algorithmic}
{\footnotesize
\STATE {\bf Initiation:} \\
\STATE \quad 1. Construct a block diagonal matrix $\D$. Perform eigenvalue decompo- \\
\STATE \quad \quad sition $\D = \U \mathbf{\Lambda} \U^{\dag}$, sort the eigenvalues in non-increasing order. \\
\STATE \quad 2. Introduce $\lambda_{0} = \infty$ and $\lambda_{K \cdot n_{t}+1} = -\infty$. Let
$\hat{I}=0$. Let the \\
\STATE \quad \quad endpoint objective value $\psi_{\hat{I}}\left(\lambda_{0}\right) = 0$, $\phi^{*}
= \psi_{\hat{I}}\left(\lambda_{0}\right)$, and $\mu^{*} = \lambda_{0}$.
\STATE {\bf Main Loop:} \\
\STATE \quad 1. If $\hat{I} > K \cdot n_{r}$, go to the final step; else let
$\mu_{\hat{I}}^{*} = (\sum_{j=1}^{\hat{I}} \lambda_{j} - P )/\hat{I}$. \\
\STATE \quad 2. If {$\mu_{\hat{I}}^{*} \in [\lambda_{\hat{I}+1}, \lambda_{\hat{I}} ] \cap
\mathbb{R}_{+}$}, then let $\mu^{*} = \mu_{\hat{I}}^{*}$ and go to the final step. \\
\STATE \quad 3. Compute $\psi_{\hat{I}}(\lambda_{\hat{I}+1})$. If $\psi_{\hat{I}}(\lambda_{\hat{I}+1}) < \phi^{*}$, then go to the final step; \\
\STATE \quad \quad else let $\mu^{*} = \lambda_{\hat{I}+1}$, $\phi^{*} = \psi_{\hat{I}}(\lambda_{\hat{I}+1})$, $\hat{I} = \hat{I} + 1$ and continue. \\
\STATE {\bf Final Step:} Compute $\tilde{\D}$ as $\tilde{\D} = \U \left( \mLambda - \mu^{*} \I
\right)_{+} \U^{\dag}$.
 }
\end{algorithmic}
\end{algorithm}

\section{Complexity Analysis} \label{sec:complexity}
In this section, we compare our proposed CGP with other existing methods for solving MIMO BC.
Similar to IWFs, SD \cite{Viswanathan03:MIMO_BC_SD}, and DD \cite{Yu_CISS03}, CGP has the desirable
linear complexity property. Although CGP also needs to compute gradients in each iteration, the
computation is much easier than that in SD due to the different perspectives in handling MIMO BC.
Thus, in this paper, we only compare CGP with IWF (Algorithms 1 and 2 in
\cite{Jindal04:MIMO_BC_IWF}), which appear to be the simplest in the literature so far. For
convenience, we will refer to Algorithm 1 and Algorithm 2 in \cite{Jindal04:MIMO_BC_IWF} as IWF1
and IWF2, respectively.

To better illustrate the comparison, we list the complexity per iteration for each component of CGP
and IWFs in Table~\ref{tab_compare}.
\begin{table}[h!]
\centering {\scriptsize \caption{\label{tab_compare}Per Iteration Complexity Comparison between CGP
and IWFs}
\begin{tabular}{|c|c|c|}
\hline  & CGP & IWFs \\
\hline Gradient/Effec. Channel & $K$ & $2K$ \\
\hline Line Search & $O(mK)$ & N/A \\
\hline Projection/Water-Filling & $O(n_{r}K)$ & $O(n_{r}K)$ \\
\hline \hline Overall & $O((m+1+n_{r})K)$ & $O((2+n_{r})K)$ \\
\hline
\end{tabular}}
\end{table}
In both CGP and IWFs, it can be seen that the most time-consuming part (increasing with respect to
$K$) is the additions of the terms in the form of $\Hc_{i}^{\dag} \Q_{i} \Hc_{i}$ when computing
gradients and effective channels. Since the term $(\I + \sum_{i=1}^{K}\Hc_{i}^{\dag} \Q_{i}
\Hc_{i})$ is common to all gradients, we only need to compute this sum once in each iteration.
Thus, the number of such additions per iteration for CGP is $K$. In IWF1 and IWF2, the number of
such additions can be reduced to $2K$ by a clever way of maintaining a running sum of $(\I +
\sum_{j \ne i}^{K}\Hc_{j}^{\dag} \Q_{j} \Hc_{j})$. But the running sum, which requires $K^{2}$
additions for IWF1, still needs to be computed in the initialization step.

Although the basic ideas of the projection in CGP and water-filling are different, the algorithm
structure of them are very similar and they have exactly the same complexity of $O(n_{r}K)$. The
only unique component in CGP is the line search step, which has the complexity of $O(mK)$ (in terms
of the additions of $\Hc_{i}^{\dag} \Q_{i} \Hc_{i}$ terms), where $m$ is the number of trials in
Armijo's Rule. Therefore, the overall complexity per iteration for CGP and IWFs are
$O((m+1+n_{r})K)$ and $O((2+n_{r})K)$, respectively. According to our computational experience, the
value of $m$ usually lies in between two and four. Thus, when $n_{r}$ is large (e.g., $n_{r} \geq
4$), the overall complexity per iteration for CGP and IWFs are comparable. However, as evidenced in
the next section, the numbers of iterations required for convergence in CGP is much less than that
in IWFs for large MIMO BC systems, and it is very insensitive to the increase of the number of
users. Moreover, CGP has a modest memory requirement: it only requires the solution information
from the previous step, as opposed to IWF1, which requires previous $K-1$ steps.

\section{Numerical Results} \label{sec:results}
\begin{figure}[ht!]
\centering
\includegraphics[width=2.8in]{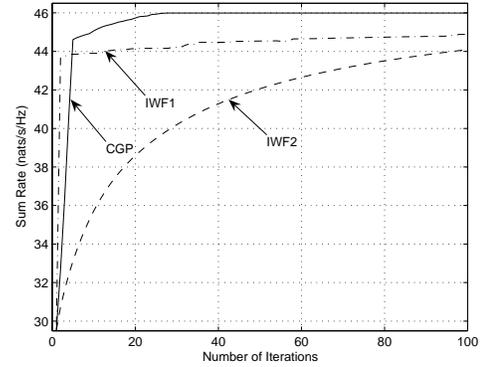}
\caption{Comparison in a 100-user MIMO BC channel with $n_{t}=n_{r}=4$.} \label{fig_example6}
\end{figure}

Due to the space limitation, we only give an example of a large MIMO BC system consisting of 100
users with $n_{t}=n_{r}=4$ in here. The convergence processes are plotted in
Fig.~\ref{fig_example6}. It is observed from Fig.~\ref{fig_example6} that CGP takes only 29
iterations to converge and it outperforms both IWFs. IWF1's convergence speed significantly drops
after the quick improvement in the early stage. It is also seen in this example that IWF2's
performance is inferior to IWF1, and this observation is in accordance with the results in
\cite{Jindal04:MIMO_BC_IWF}. Both IWF1 and IWF2 fail to converge within 100 iterations. The
scalability problem of both IWFs is not surprising because in both IWFs, the most recently updated
covariance matrices only account for a fraction of $1/K$ in the effective channels' computation,
which means it does not effectively make use of the most recent solution. In all of our numerical
examples with different number of users, CGP always converges within 30 iterations.

\section{Conclusion} \label{sec:conclusions}
In this paper, we developed an efficient algorithm based on conjugate gradient projection (CGP) for
solving the maximum sum rate problem of MIMO BC. We theoretically and numerically analyzed its its
complexity and convergence behavior. The attractive features of CGP and encouraging results showed
that CGP is an excellent method for solving the maximum sum rate problem for large MIMO BC systems.

\end{document}